\documentclass[runningheads]{llncs}
\usepackage[T1]{fontenc}
\usepackage{graphicx}
\usepackage{hyperref}
\usepackage{color}

\usepackage[frozencache,cachedir=minted-cache]{minted2}
\usepackage{amssymb}
\usepackage{tikz}
\usetikzlibrary{positioning}
\tikzset{
    node base/.style = {rectangle, rounded corners, draw=black, thick},
    arrow base/.style = {thick}
}
\usepackage{algorithmic}
\usepackage{algorithm}
\usepackage{booktabs}
\begin{document}
\title{Learning Rules Explaining Interactive Theorem Proving Tactic Prediction}

\titlerunning{Learning Rules Explaining Interactive Theorem Proving Tactic Prediction}
%
\author{Liao Zhang\inst{1}\orcidID{0000-0002-4574-8843} \and
David M. Cerna\inst{2}\orcidID{0000-0002-6352-603X} \and
Cezary Kaliszyk\inst{3,1}\orcidID{0000-0002-8273-6059}}
\authorrunning{L. Zhang et al.}
%
\institute{University of Innsbruck, Innsbruck, Austria \\
\email{Liao.Zhang@student.uibk.ac.at} \and
Czech Academy of Sciences, Prague, Czechia\\
\email{dcerna@cas.cs.cz}  \and
University of Melbourne \\
\email{cezary.kaliszyk@unimelb.edu.au} 
}
\maketitle              
\begin{abstract}
Formally verifying the correctness of mathematical proofs is more accessible than ever, however, the learning curve remains steep for many of the state-of-the-art interactive theorem provers (ITP). Deriving the most appropriate subsequent proof step, and reasoning about it, given the multitude of possibilities, remains a daunting task for novice users. To improve the situation, several investigations have developed machine learning based guidance for \textit{tactic} selection.
Such approaches struggle to learn non-trivial relationships between the chosen tactic and the structure of the proof state and represent them as symbolic expressions.

To address these issues we (i) We represent the problem as an \emph{Inductive Logic Programming (ILP)} task, (ii) Using the ILP representation we enriched the feature space by encoding additional, computationally expensive properties as \emph{background knowledge} predicates, (iii) We use this enriched feature space to learn rules explaining when a tactic is applicable to a given proof state, (iv) We use the learned rules to filter the output of an existing tactic selection approach and empirically show improvement over the non-filtering approaches.

\end{abstract}

\section{Introduction\label{sec:intro}}
Interactive Theorem Provers (ITP), such as Coq~\cite{coq811}, Lean~\cite{de2015lean}, and Isabelle~\cite{paulson1994isabelle}, are powerful tools that combine human instruction with computer verification to construct formal mathematical proofs, providing a reliable means of certification and ensuring safety in critical applications. 

These systems operate as follows: the user specifies a goal to prove,  \emph{the initial proof state}. Then the user specifies \textit{tactics} (an operation transforming a proof state into proof states). Certain tactics close proof states. The proof is complete if there are no remaining open proof states, i.e., the goal has been proved. 

Given the complexity of ITP systems, a fully automated approach to proving user specified goals is intractable. Numerous investigations have instead focused on providing the user with guidance through tactic suggestion. 

The methods used in practice by ITP users are statistical machine learning methods such as $k$-nearest neighbors ($k$-NN) and naive Bayes~\cite{gauthier2021tactictoe}. These methods take a goal $g$, select a goal $g'$ most similar goal to $g$, and rank the particular tactics relevant for solving $g'$ based on their likelihood of solving $g$.  

Neural network and LLM-based approaches addressing the task include:
CoqGym~\cite{yang2019learning} trains tree neural networks to automatically construct proofs for \textit{Coq}.
Thor~\cite{jiang2022thor} combines LLMs and external symbolic solvers to search for proofs for Isabelle.
LLMs are also applied to synthesising training data to enhance the performance of theorem proving~\cite{xin2023lego}. 
Despite showing slight improvement in performance during  machine learning evaluations, in practice these methods require long training for each new theory, which makes them less useful for day to day proof development. 

Additionally, they lack interpretability. When a user receives predictions, they may want to know why a particular tactic was chosen over another tactic to better understand what actions they should take in the future. 

Furthermore, guidance based on statistical learning approaches often requires propositionalisation of features, calculated based on the structure of the \emph{abstract syntax tree (AST)} of a proof state~\cite{zhang2021online}, e.g., \textit{there is a path between nodes X and Y in tree T}. For complex and precise features, pre-computation is prohibitively expensive.

Moreover, logical inference is significantly influenced by the small error margins present in the statistical inferencing mechanisms of LLMs and similar models. Thus, predictions based on chained logical inferences will quickly suffer a loss of predicative accuracy~\cite{lecun2023large}. 

In contrast to pre-computed features, we represent such features as logic programs and compute them only when needed for learning. For example, we define logic programs for the existence of two particular nodes on a path (of arbitrary length) from the root of the tree as ($\mathit{above}(AST,X,Y)$). Below, we present a learned rule for the simplification tactic which states that the tactic is applicable to a proof state when the goal node of the proof state contains a constant above two constructs (also in the goal) which differ. 
 
\begin{minted}{prolog}
tac(A,"simpl") :-
  goal_node(const,A,B,C), goal_node(construct,A,D,E),
  goal_above(A,B,D), goal_node(construct,A,F,E),dif(F,D), 
  goal_above(A,B,F).     
\end{minted}

The rules, as presented above, are learned using inductive logic programming (ILP), in particular, \textit{Aleph}~\cite{srinivasan2001aleph}. In addition to providing rules explaining tactic prediction, we use the resulting rules to filter the output of $k$-NN, in particular, the classifier presented in~\cite{blaauwbroek2020tactician,gauthier2021tactictoe} (Tactician and TacticToe). Essentially, we want to determine whether $ps,r\vDash p_t$ where $ps$ is a logic program representing the proof state, $r$ is a learned rule for the tactic $t$, and $p_t$ is the head predicate of $r$ denoting that $t$ should be applied to $ps$. Thus, given the list of recommended tactics by a $k$-NN classifier, we can further filter this list using the learned rules. Our hypothesis is that features of proof state defined through logic programs can be used to learn rules which can be used to filter the output of a  $k$-NN model to improve accuracy.

In addition to improved performance, our approach produces rules to explain the predictions.
Consider again the aforementioned rule of \texttt{simpl}  that specifies that the goal may be simplified if it contains a constant above two constructors with different positions. Here, the constructor and the constant denote the datatypes of Coq's terms. 
The same variable $E$ confirms that the two constructors must correspond to the same identifier in Coq. This rule may suit the Coq structure $S \ x - S \ y$ which denotes $(1+x)-(1+y)$.
It can be simplified to $x - y$.
$S$ denotes a constructor, and $-$ denotes a constant.
The first argument of \texttt{goal\_node} is a constant that is constrained by us via mode declarations~\cite{srinivasan2001aleph}. 

We use the ILP system Aleph~\cite{srinivasan2001aleph} together with a user-defined cost function to evaluate the learned rules on Coq's standard library.
We chose Aleph because it has empirically good results~\cite{cropper2022inductive}. We refrain from using modern ILP approaches such as \textit{Popper}~\cite{cropper2021learning} as the underlying ASP solvers have difficulty generating models when many variables are required and high-arity definitions are included in the background.
We develop representation predicates (\texttt{goal\_node}) to efficiently denote the nodes of the AST.
We also develop feature predicates (\texttt{goal\_above}) which denote the properties of the AST calculated based on the representation predicates. The motivation for developing feature predicates is that propositionalization of it would significantly enlarge the representation making it impractical to use.
Our experiments confirm that feature predicates can learn more precise rules (rules with higher F-1 scores~\cite{sasaki2007truth}) compared to representation predicates. 
Additionally, the experiments demonstrate that the combination of ILP and $k$-NN can improve the accuracy of tactic suggestions in Tactician, the main tactic prediction system for Coq.

\paragraph{Contributions}
First, we express the task of predicting the best tactic to apply to the given proof state as an ILP task.
Second, using the ILP representation we enriched the feature space by encoding additional, computationally expensive features as \emph{background knowledge} predicates, allowing us to avoid grounding the features which are computationally expensive. 
Third, We use this enriched feature space to learn rules explaining when a tactic is applicable to a given proof state and filter the output of an existing tactic selection approach using these rules.
Finally, We empirically show improvement over the non-filtering approaches.

This is the first time an investigation has considered ILP as a tool for improving tactic suggestion methods for ITPs.

\section{Background~\label{sec:background}}
\begin{figure}[t]
    \centering
    \includegraphics[width=\textwidth]{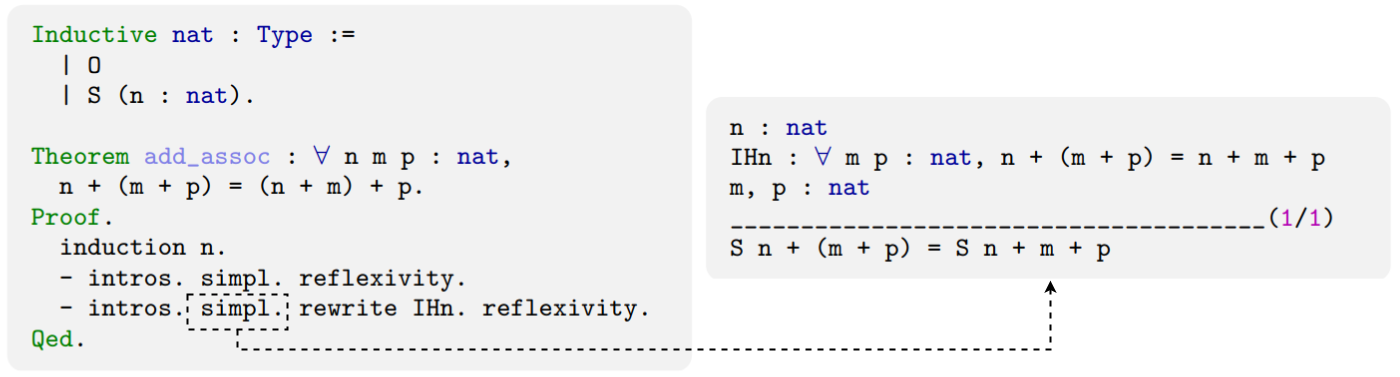}
    \caption{A Coq proof of the associative property of addition and the proof state before \texttt{simpl}.}
    \label{fig:coq-proof}
\end{figure}

\paragraph{Theorem Proving in Coq}
Coq is one of the most popular proof assistants and has been widely used for building trustworthy software~\cite{leroy2021compcert} and verifying the correctness of mathematical proofs~\cite{gonthier2008formal}.
Coq tactics are proof state transformations that provide a high-level combination of underlying logical inferences.
To illustrate how theorems are formalized in Coq, we present a simple example in Figure~\ref{fig:coq-proof}. Here, we want to prove the associative property of addition. The natural numbers in Coq are defined by two constructors \texttt{O} and \texttt{S}. \texttt{O} denotes 0, and \texttt{S n} denotes $n + 1$.
Here, the initial proof state is the same as the statement of the theorem.
We first apply induction on \texttt{n} and obtain two cases corresponding to the two constructors. In the first case, \texttt{n} equals 0. After some simplifications, we can prove $0 = 0$ by the tactic \texttt{reflexivity}. The second case is a bit more complicated, and we need to apply the induction hypothesis \texttt{IHn} to finish the proof.
Figure~\ref{fig:coq-proof} also presents a concrete example of a proof state. A proof state consists of a \textit{goal} to prove and several \textit{hypotheses}. The goal is below the dashed line. \texttt{IHn}, \texttt{n}, \texttt{m}, and \texttt{p} are the names of hypotheses. A proof state is often represented as a sequent $E \vdash g$ where $E$ and $g$ denote the hypotheses and the goal, respectively.

\paragraph{Adaptations of $k$-NN to Theorem Proving}
Several machine learning algorithms have been adapted to theorem proving tasks. In most cases, simpler algorithms adapted to formal reasoning tasks perform better than deep learning based methods in practice. For this reason, modified $k$-NN (explained in~\cite{blaauwbroek2020tactic}) is the main algorithm for TacticToe and Tactician.
Even if evaluations with deep learning or large tree-based classifiers have shown some theoretical improvements, the simpler algorithms training for the particular theories developed by users, give a larger practical advantage to the users. As such, we focus on the modified $k$-NN in this work. 
Standard $k$-NN starts by calculating the distance between a new proof state and all known proof states in the database.
The distance is measured by the similarity between the features of the proof states, usually using tree walks in the AST of the proof state~\cite{ckjujv-ijcai15}. The dependencies of such selected neighbours, with additional scaling by their distances, inclusion of the neighbours themselves, and further modifications exhibit commendable empirical performance~\cite{rute2024graph2tac}, and are therefore the default algorithm both in Tactictoe and Tactician.

\section{Background Knowledge}
To utilize ILP, we need to appropriately define the background knowledge.
We start this section by encoding the nodes of AST as representation predicates.
Then we propose new feature predicates that will allow leveraging the power of ILP. We finally add predicate anonymization, already very useful in automated reasoning systems, to representation and feature predicates.

\subsection{Representation Predicates}
Every node in the AST of the proof state is converted to a fact.
There are two categories of nodes: identifiers of existing objects and constructors of Coq's datatype.
A node in the goal is converted to  $goal\_node(name, nat, goal\_idx)$. The argument $name$ refers to the value of the node.
A unique natural number is assigned to every proof state to identify it. The argument $goal\_idx$ uses a sequence of natural numbers to specify the position of the node in the goal. 
A node in a hypothesis is converted to a fact $hyp\_node(name, nat, hyp\_name, hyp\_idx)$.
Compared to a $goal\_idx$, a $hyp\_idx$ starts with the name of a hypothesis so that two $hyp\_idx$ from different hypotheses have different prefixes.
The $goal\_node$ and $hyp\_node$ predicates are called \textit{representation predicates} in this work. 

\subsection{Feature Predicates\label{sec:rel-predc}}
We also define two categories of \textit{feature predicates} which represent the properties of AST based on the representation predicates.
\paragraph{Positional Predicates} represents the relative relationships between nodes' positions. 
The predicate \texttt{goal\_left(Goal\_idx1, Goal\_idx2)} and \texttt{goal\_above(Nat, Goal\_idx1, Goal\_idx2)} respectively checks whether the node is left (above) to another node in the goal.
They are inspired by the horizontal features and vertical features used in previous works~\cite{chvalovsky2019enigma,zhang2021online}. 
Similarly, we define \texttt{hyp\_left(Hyp\_idx1, Hyp\_idx2)} and \texttt{hyp\_above(Nat, Hyp\_idx1, Hyp\_idx2)}. 
Previous works have confirmed the usefulness of using the occurrence numbers of features in feature characterization, 
which inspires us to develop the predicate \texttt{dif(Goal\_idx1, Goal\_idx2)}.
It denotes that the same node multiply occurs in different positions in the goal.

\paragraph{Equational Predicates} check the equality between two terms. 
The predicate \texttt{eq\_goal\_term(Nat, Goal\_idx1, Goal\_idx2)} checks that the two subterms in the goal are the same. The root nodes of the two subterms are located in the positions \texttt{Goal\_idx1} and \texttt{Goal\_idx2}, respectively.
It pertains to \texttt{reflexivity} which proves a goal of the equation if the equality holds after some normalization. 
Thus, it can prove $x = x$ and inspires us to develop \texttt{eq\_goal\_term}.
The predicate \texttt{eq\_goal\_hyp\_term(Nat, Goal\_idx, Hyp\_idx)} is inspired by a number of tactics that check the equality between the goal and the hypotheses, such as \texttt{assumption}, \texttt{apply}, and \texttt{auto}. 
For instance, \texttt{assumption} proves a goal if it equals a hypothesis.
Assume a proof state $H_1 : Q \ x, H_2 : P \ x \rightarrow Q \ x \vdash Q \ x$ which can be proved by \texttt{assumption}.
The predicate \texttt{eq\_goal\_hyp\_term} checks the equality between the goal and $Q \ x$ in a hypothesis.
The predicate \texttt{is\_hyp\_root(Nat, Hyp\_idx)} ensures the node is the root of a hypothesis. Thus, it can show the equality only holds between the goal and $H_1$ instead of $H_2$.
With a reason akin to that of \texttt{is\_hyp\_root}, we define \texttt{is\_goal\_root(Nat, Goal\_idx)}.
The equality between two terms in different hypotheses is checked by \texttt{eq\_hyp\_term(Nat, Hyp\_idx1, Hyp\_idx2)}.
It is useful for tactics that can apply hypotheses several times, e.g., \texttt{auto}.
Assume a proof state $H_1 : P \ x, H_2 : P \ x \rightarrow Q \ x \vdash Q \ x$. First, \texttt{auto} applies $H_2$ to the goal and changes the goal to $P \ x$. Then, it applies $H_1$ to prove the new goal. The description of the operation requires to show that $H_1$ equals to the premise of $H_2$.

\subsection{Anonymous Predicates}
We also substitute identifiers with more abstract descriptions to facilitate the generalization ability of ILP.
The substitution is similar to that in ENIGMA anonymous~\cite{jakubuuv2020enigma}.
The predicates that accept original nodes and abstract nodes as their first arguments are called \textit{original predicates} and \textit{anonymous predicates}, respectively.
We substitute identifiers with their categories, consisting of inductive types, constants, constructors, and variables.
Besides the abstract nodes, we also include the original nodes as arguments in \texttt{goal\_node} and \texttt{hyp\_node}.
We need them because when checking the equality, we want to compare the original nodes. 
Afterward, the anonymous predicates of nodes change to\\ $goal\_node(anonym\_name, nat, goal\_idx, origin\_name)$ and\\ $hyp\_node(anonym\_name, nat, hyp\_name, hyp\_idx, origin\_name)$.
Some basic identifiers are not substituted, which consist of  \texttt{logic\_false}, \texttt{logic\_true}, \texttt{and}, \texttt{or}, \texttt{iff},  \texttt{not}, \texttt{eq}, \texttt{bool\_true}, and \texttt{bool\_false}.
There are both logic and boolean values of true and false because Coq can represent objects in logic or programs. 
Concerning the constructors of Coq's datatypes, we only retain four important constructors: \texttt{rel}, \texttt{prod}, \texttt{lambda}, and \texttt{evar}.

\section{Method\label{sec:method}}
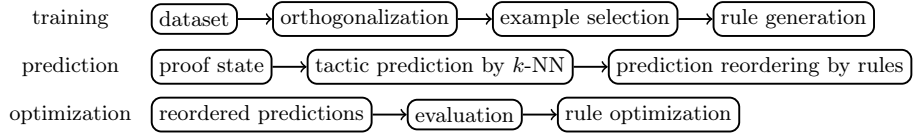
\begin{figure*}[t]
\centering
\resizebox{\textwidth}{!}{
\begin{tikzpicture}[node distance=7mm and 5mm]
    \node (a0) {training};
    \node (a1) [node base, right=of a0] {dataset};
    \node (a2) [node base, right=of a1] {orthogonalization};
    \node (a3) [node base, right=of a2] {example selection};
    \node (a4) [node base, right=of a3] {rule generation};

    \node (b0) [below =of a0.north] {prediction};
    \node (b1) [node base, below =of a1.west, anchor=west] {proof state};
    \node (b2) [node base, right =of b1] {tactic prediction by $k$-NN};
    \node (b3) [node base, right =of b2] {prediction reordering by rules};

    \node (c0) [below=of b0.north] {optimization};
    \node (c1) [node base, below=of b1.west, anchor=west] {reordered predictions};
    \node (c2) [node base, right =of c1] {evaluation};
    \node (c3) [node base, right =of c2] {rule optimization};
    
    \path[arrow base, ->] (a1) edge node {} (a2);
    \path[arrow base, ->] (a2) edge node {} (a3);
    \path[arrow base, ->] (a3) edge node {} (a4);

    \path[arrow base, ->] (b1) edge node {} (b2);
    \path[arrow base, ->] (b2) edge node {} (b3);

    \path[arrow base, ->] (c1) edge node {} (c2);
    \path[arrow base, ->] (c2) edge node {} (c3);
\end{tikzpicture}}
\caption{An overview of the procedures of the learning framework.}
\label{fig:overview}
\end{figure*}
Figure~\ref{fig:overview} presents an overview of our learning framework. During the training, we first perform orthogonalization, a technique introduced in TacticToe, to clean the dataset. Then, we select examples and apply ILP to generate rules. To make predictions, first, $k$-NN predicts a sequence of likely helpful tactics. Afterward, the rules are used as a filter to reorder the predictions. The optimization procedure denotes removing some low-quality rules. This is achieved by evaluating the reordered predictions in the validation dataset and removing the low-quality ones. In the next subsections, we describe these parts.

\subsection{Orthogonalization~\label{sec:ortho}}
In some cases, different tactics could transform the same proof state in the same way. This raises ambiguity and makes learning difficult. Orthogonalization is used to reduce such ambiguity. 
In the orthogonalization, we only focus on four very popular tactics in Coq's standard library: \texttt{assumption}, \texttt{reflexivity}, \texttt{trivial}, and \texttt{auto}. 
We denote the sets of proof states which can be closed by \texttt{assumption}, \texttt{reflexivity}, \texttt{trivial}, and \texttt{auto} as $AS$, $R$, $T$, and $AT$, respectively.
There exist the relations $AS \subsetneq T$, $R \subsetneq T$, and $T \subsetneq AT$.
For each proof state $ps$ to which the tactic $t$ is applied, the above four automation tactics are sequentially tried.
If $ps$ can be finished by the automation tactic $t'$, we replace $t$ by $t'$.
If none of the four tactics can finish the proof state, the original $t$ is preserved.
The orthogonalization procedure is simpler than in TacticToe, which orthogonalizes all tactics. This is
because our current predicates can only capture a part of the usage of tactics.
We leave full orthogonalization as future work.

\subsection{Example Selection~\label{sec:exg_selection}}
Choosing appropriate training examples is crucial for learning reasonable rules.
For a specific tactic $tac$, the proof states to which it is applied are regarded as the positive examples.
The proof states to which the tactics different from $tac$ are applied are regarded as the negative examples. 
We experimentally determine the number of positive and negative examples for learning rules.
We develop a clustering mechanism to split positive examples into roughly equal-sized clusters. 
We experimentally evaluate the combinations of different numbers of negative examples and different numbers of positive examples. 

We choose an implementation of a constrained $k$-means algorithm~\cite{Levy-Kramer_k-means-constrained_2018} to split positive examples into clusters of roughly the same size. The original $k$-means algorithm~\cite{hartigan1979algorithm} can only split examples into a certain number of clusters. In contrast, constrained $k$-means can also specify the lower bound and the upper bound of the size of the clusters, which is important to give good sizes of training examples for each ILP learning task.

We apply $k$-NN to discover negative examples for each positive example. As this pre-processing step is not theorem-proving specific, 
we use the general $k$-NN from the scikit-learn library~\cite{scikit-learn}.
We use the same features as Tactician~\cite{zhang2021online}.
For each positive example, $k$-NN calculates the distance between it and every negative example in the training data. Then, we rank the negative examples in an ascending order of distance.

\subsection{Training and Prediction\label{sec:reorder}}
For each tactic, we use Aleph to generate ILP rules for each cluster of positive examples and its associated negative examples.
Afterwards, all the rules are merged together, and duplicated rules are removed.
Finally, we remove the rules of tactics that are logically subsumed by other rules of the same tactic.

Algorithm~\ref{alg:reorder} illustrates the procedures of making predictions. 
We use the state-of-the-art $k$-NN in Tactician.
The features are the same as those used in Section~\ref{sec:exg_selection}.
Assume a pair of a proof state and a tactic $(ps, tac)$. To make predictions, first, $k$-NN preselects a sequence of likely tactics $tac_{1..50}$. For each $tac_i$, we use the learned rules to determine whether to accept it.
During the evaluation, the prediction $tac_i$ is expected (unexpected) if $tac_i$ is equal (unequal) to $tac$. 
If the rules accept (reject) a tactic, the prediction is a declared positive (negative). 
If the rules reject a $tac_i$ equal to $tac$, we regard the prediction made by the rules as a false negative (FN).
Based on the expected tactics and acceptances, we also obtain true positives (TPs), true negatives (TNs), and false positives (FPs).

\begin{algorithm}[t]
\caption{Preselection Reorder}\label{alg:reorder}
\begin{algorithmic}
\STATE {\bfseries Input:} a sequence of tactics $tac_{1..50}$ preselected by $k$-NN for a proof state
\STATE {\bfseries Output:} a sequence of tactics which is a reorder of the preselection 
\STATE $goods \gets $ [ ]
\STATE $bads \gets $ [ ]
\FORALL{$i \in \{1..50\}$ }
    \IF{$tac_i$ is accepted by learned rules} 
        \STATE append $tac_i$ to the end of $goods$
    \ELSE
        \STATE append $tac_i$ to the end of $bads$
    \ENDIF
\ENDFOR
\STATE $reorder \gets$ the sequence of appending $bads$ to the end of $goods$
\STATE {\bfseries return} $reorder$
\end{algorithmic}
\end{algorithm}

\subsection{Rule Optimization~\label{sec:rule_optim}}
The idea of rule optimization is to remove some low-quality rules to increase the overall performance of rules.
For the evaluation of all rules, we chose the F-1 score as the metric, defined as $\frac{2TP}{2TP+FP+FN}$, because it is a standard metric for evaluating imbalanced data.
As an illustration of the imbalance, given a pair of a proof state and a tactic $tac$, rules make predictions for 50 preselected tactics. However, at most one is the same as $tac$.
If a rule is overly general, which means that the number of FPs introduced by it is much larger than the number of TPs introduced by it, removing it will increase the overall F-1 score.

Although a large number of negative examples prevents generating overly general rules, using them may not produce the best rules for two reasons.
First, our background can merely capture a portion of the usage of the tactics; thus, a significantly large number of negative examples cannot produce perfect rules but may produce overly specific rules.
Second, some negative examples in our dataset are actually false negatives.
A mathematician may be able to choose the next step from a couple of tactics that make different proof transformations.
Orthogonalization in Section~\ref{sec:ortho} can only partially remove such overlaps between tactics, thereby decreasing the number of false negatives. It is computationally prohibitive to perform full orthogonalization of our data. Observe that, our experiments still show an increase in accuracy in light of the noisy data.

Our approach allows us to learn many rules explaining a particular tactic. Over the training set, some of these rules capture the usage of the given tactic better than others. Before, moving to testing on unseen data, we prune the learned rules and keep only those that performed well on the validation set.

To determine which rules to include, we evaluate the quality of each rule in the validation dataset and remove those with low qualities.
The left rules are used for the evaluation on the test dataset.
We measure the quality of each rule and remove it if its quality is below a certain threshold. 
We set different thresholds and choose the threshold leading to the highest F-1 score via experiments.
For the metric of the quality of a single rule, we use \textit{precision}, defined as $\frac{TP}{TP+FP}$.
Here, $FP$ and $TP$ are produced by a single rule.
Precision is a good metric because if a rule is too general, its precision will be low and we will be able to remove it to improve the overall F-1 score.

\section{Experiments}
We conducted the experiments on Coq's standard library~\cite{czajka2018hammer}.
We chose Coq's standard library as the benchmark because it is a standard dataset for evaluating machine learning for Coq.
Moreover, it comprises well-crafted proofs developed by Coq experts and has been optimized for decades.
Coq's standard library consists of 41 theories and 151,678 proof states in total. The code for this paper is available at \url{https://github.com/Zhang-Liao/ilp_coq}.

Most parameters of Aleph were left as default besides three parameters.
We set the maximal length of a clause to 1,000, the upper bound of proof depth to 1,000, and the largest number of nodes to be explored during the search to 30,000.
We define a cost function similar to the default cost function because, by default, Aleph cannot learn with no negative examples or only one positive example.
The user-defined cost function was only used when there were no negative examples or exactly one positive example.
We set a timeout of ten minutes for learning.

We conducted the experiments in the transfer-theory setting, which means different Coq theories are used for training, validation, and testing.
We use this setting because it simulates a practical application scenario of ILP.
Mathematicians develop new theories based on the definitions and proven theorems in the developed theories.
To be practically beneficial, ILP should also learn rules from training theories, and the learned rules should help make tactic suggestions for theories that do not depend on the training theories.  
The training theory should be carefully chosen before conducting experiments.
The theory \texttt{Structures} was chosen for training because it has a balanced distribution of various tactics. 
To be consistent with the transfer-theory setting, the testing theories should not depend on \texttt{Structures}.
From Coq's standard library, we chose all theories which do not depend on \texttt{Structures} for testing including \texttt{rtauto}, \texttt{FSets}, \texttt{Wellfounded}, \texttt{funind}, \texttt{btauto}, \texttt{nsatz}, and \texttt{MSets}.
Afterward, from all the theories that do not depend on the testing theories, we randomly chose five theories: \texttt{PArith}, \texttt{Relations}, \texttt{Bool}, \texttt{Logic}, and \texttt{Lists}, merged as the validation dataset.

\begin{figure*}[t]
\centering
\includegraphics[width=\textwidth]{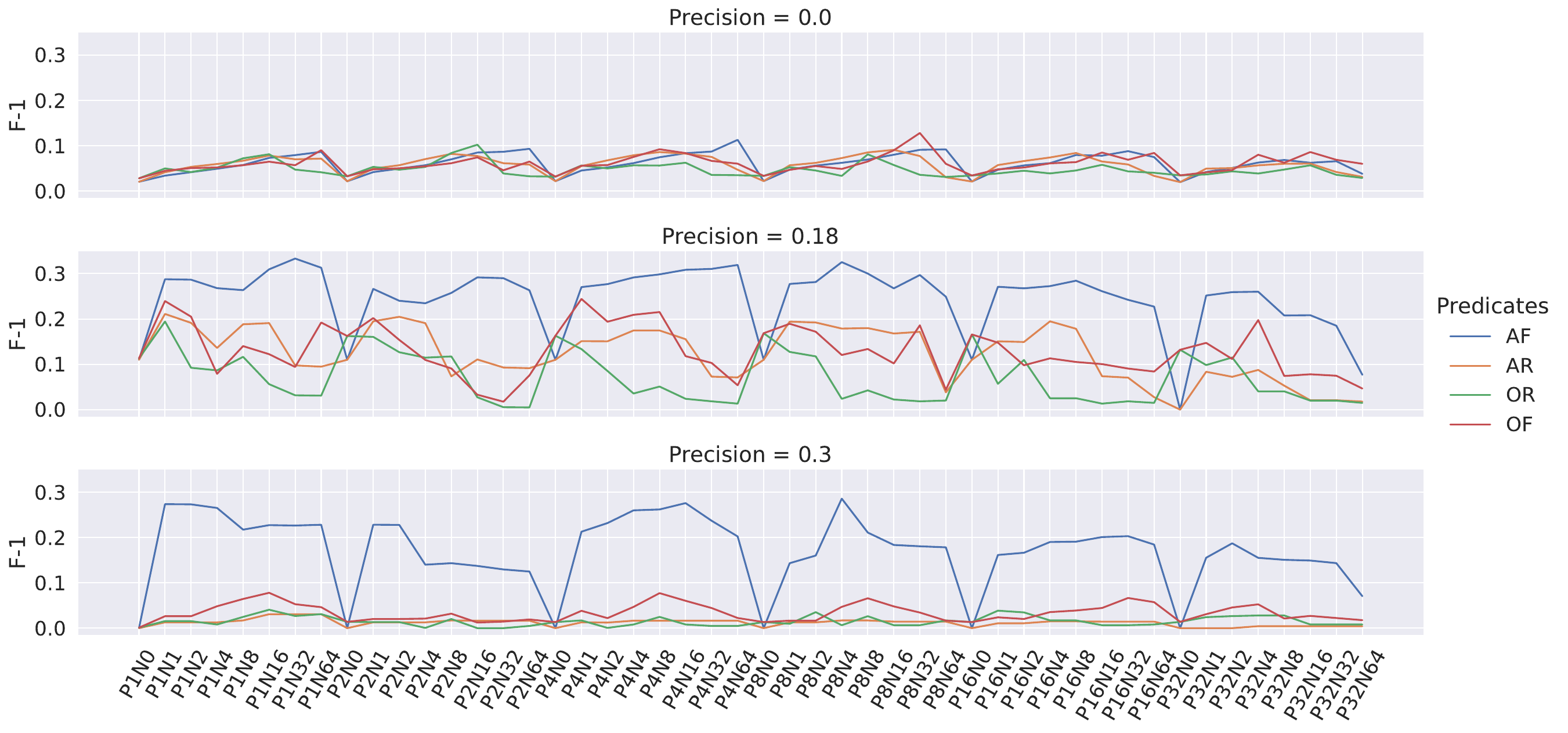}
\caption{F-1 scores of different parameters when $qualt$ is set to 0, 0.18, or 0.30. \textit{AF}, \textit{AR}, \textit{OF}, and \textit{OR} denote the anonymous feature predicates, the anonymous representation predicates, the original feature predicates, and the original representation predicates, respectively. In the x-axis caption \textit{P} and \textit{N} denote $pos$ and $neg$, respectively.}
\label{fig:tune}
\end{figure*}

\subsection{Parameter Optimization\label{sec:param_optim}}
In Section~\ref{sec:method}, we introduced three additional hyper-parameters beyond those already present in Aleph.
They are the size of the cluster of positive examples ($pos$), the number of negative examples of each positive example ($neg$), and the quality-theshold ($qualt$) below which the rule should be removed.

We evaluated the F-1 scores of different predicate categories with different parameters.
There are four predicate categories AF, AR, OF, and OR, respectively denoting the anonymous feature predicates, the anonymous representation predicates, the original feature predicates, and the original representation predicates. $AF$ and $OF$ contain both representation predicates and feature predicates, while $AF$ and $AR$ only contain anonymous predicates.
We chose $pos$ between 0 and 32. 
For $neg$, we chose it between 0 and 64.
For all the combinations of $pos$ and $neg$, rules were generated.
Afterward, the learned rules were evaluated in the validation dataset.
Finally, we calculated the F-1 scores with different values of $qualt$.
The range of $qualt$ was set between 0 and 0.30, with intervals of 0.06.

Figure~\ref{fig:tune} depicts the F-1 scores when $qualt = \{0, 0.18, 0.30\}$.
The significance of $qualt$ is evident.
When $qualt=0$, the best F-1 scores of all predicate categories hover around 0.10.
The best scores become significantly higher than 0.10 when $qualt=0.18$.
The low F-1 scores of $qualt = 0$ are caused by some overly general rules which are discussed in Section~\ref{sec:rule_optim}.
An example of such an overly general rule is provided below, showing the necessity of employing an appropriate $qualt$.
\begin{minted}{prolog}
tac(A,"reflexivity") :- goal_node(coq_Init_Logic_eq,A,B,C).    
\end{minted}
The above rule denotes that \texttt{reflexivity} is appropriate whenever there is an equal sign in the goal.
It is too general and irrelevant to the usage of \texttt{reflexivity} as explained in Section~\ref{sec:rel-predc}.
With $qualt=0.30$, the best F-1 scores decrease again. 
The decline is attributed to the fact that most rules remained by a very high $qualt$ are excessively specific, thereby producing a limited number of TPs.

Afterward, we analyze the results obtained with $qualt=0.18$ since the F-1 scores are notably higher than those with $qualt = \{0, 0.30\}$.
None of the predicate categories obtains the highest F-1 score with $pos = 32$.
We assume the reason is that an overly large $pos$ may gather many irrelevant positive examples, which causes difficulties in
choosing negative examples. 
If two positive examples significantly differ, an appropriate negative example for one of them may be inappropriate for the other.
A large value of $neg$ generally decreases the F-1 scores of all predicates except for $AF$.
A possible explanation is that too many negative examples cause $AR$, $OF$, and $OR$ to learn overly specific rules.
Due to the expressivity of $AF$, it can still learn some reasonable rules.

Table~\ref{tab:best-param} displays the optimal parameters of all predicate categories.
The generalization does not work well for $OF$ and $OR$, but already
with $AF$ its F-1 score peak necessitates a large $neg$, indicating its superior ability to distinguish positive examples from negative examples and to learn precise rules.
Perhaps due to the reason that our background knowledge is incapable of perfectly capturing the usage of tactics, $AF$ also uses $pos = 1$.
A small $pos$ allows $AF$ to learn many rules for diverse situations.
$AR$ requires $pos = 16$ to achieve its peak F-1 score, possibly due to its limitation of representing AST in a highly generalized manner.

\begin{table}[t]
  \begin{minipage}{0.35\textwidth}
    \caption{The best parameters of each predicate category.}
    \label{tab:best-param}
    \centering
      \begin{small}
        \begin{sc}
          \begin{tabular}{|l|c|c|c|c|}
           \hline
            Parameter & AF   & AR  & OF  & OR  \\
            \hline
            Precision & 0.18 & 0.12 & 0.18 & 0.12 \\
            Positive  & 1    & 16   & 4   & 1   \\
            Negative  & 32   & 1   & 1   & 1   \\
          \hline
          
          \end{tabular}
        \end{sc}
      \end{small}
  \end{minipage}
  \begin{minipage}{0.65\textwidth}
    \caption{The F-1 scores in the test dataset.}
    \label{tab:test_f1}
    \centering
      \begin{small}
        \begin{sc}
          \begin{tabular}{|l|c|c|c|c|}
            \hline
            Theory      & AF             & AR    & OF    & OR    \\
            \hline
            rtauto      & \textbf{0.564} & 0.401 & 0.502 & 0.440 \\
            FSets       & \textbf{0.266} & 0.125 & 0.193 & 0.144 \\
            Wellfounded & \textbf{0.229} & 0.049 & 0.134 & 0.135 \\
            funind      & \textbf{0.545} & 0.0   & 0.0   & 0.0 \\
            btauto      & \textbf{0.339} & 0.125 & 0.162 & 0.122 \\
            nsatz       & \textbf{0.164} & 0.070 & 0.163 & 0.116 \\
            MSets       & \textbf{0.272} & 0.084 & 0.143 & 0.095 \\
           \hline
          \end{tabular}
        \end{sc}
      \end{small}
  \end{minipage}
\end{table}

\begin{figure*}[t]
\centering
\centerline{\includegraphics[width=\textwidth]{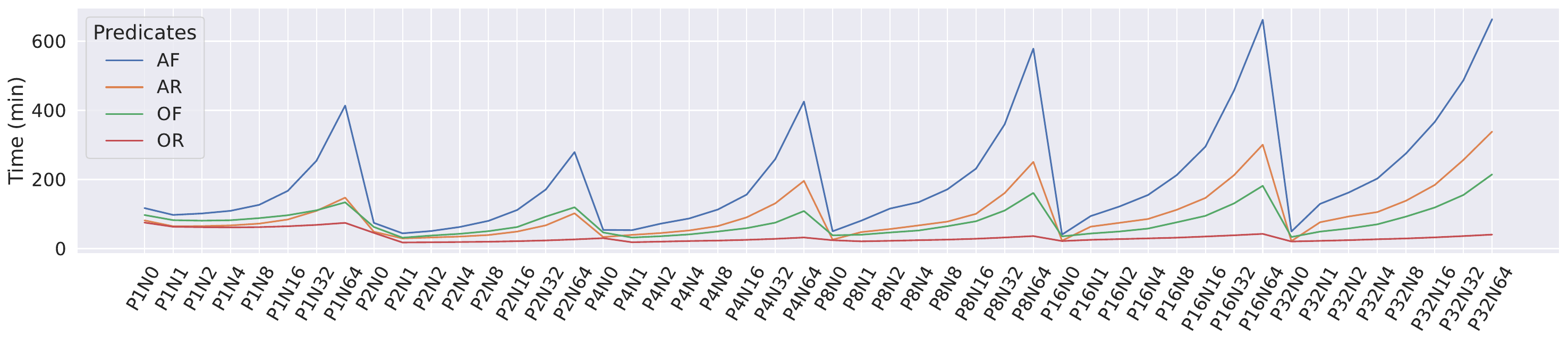}}
\caption{Training times for different categories of predicates.}
\label{fig:time}
\end{figure*}

The training times for each category of predicates are presented in Figure~\ref{fig:time}.
The total training time is calculated by summing the training times for the examples associated with each cluster of  tactics.
Learning definitions of feature predicates is computationally more costly than learning definitions of representation predicates, likely due to the resource overhead required for the former.
Moreover, learning anonymous predicates requires more time than original predicates.
We hypothesize that obfuscating some of the structure of the proof terms (when learning  anonymous predicates) correlates with an increase in the size of the search space and thus longer learning times when searching for programs distinguishing a large number of positive examples from negative examples.
In some cases the learning phase took more than ten hours, indicating potential future work to improve the computational efficiency. 
Nevertheless, our learning is still significantly faster than contemporary neural networks trained to solve similar problems and, in addition, we provide an explanation of our decisions (in the form of a logic program).

\subsection{Testing\label{sec:test}}
\begin{figure*}[t]
\centering
\centerline{\includegraphics[width=\textwidth]{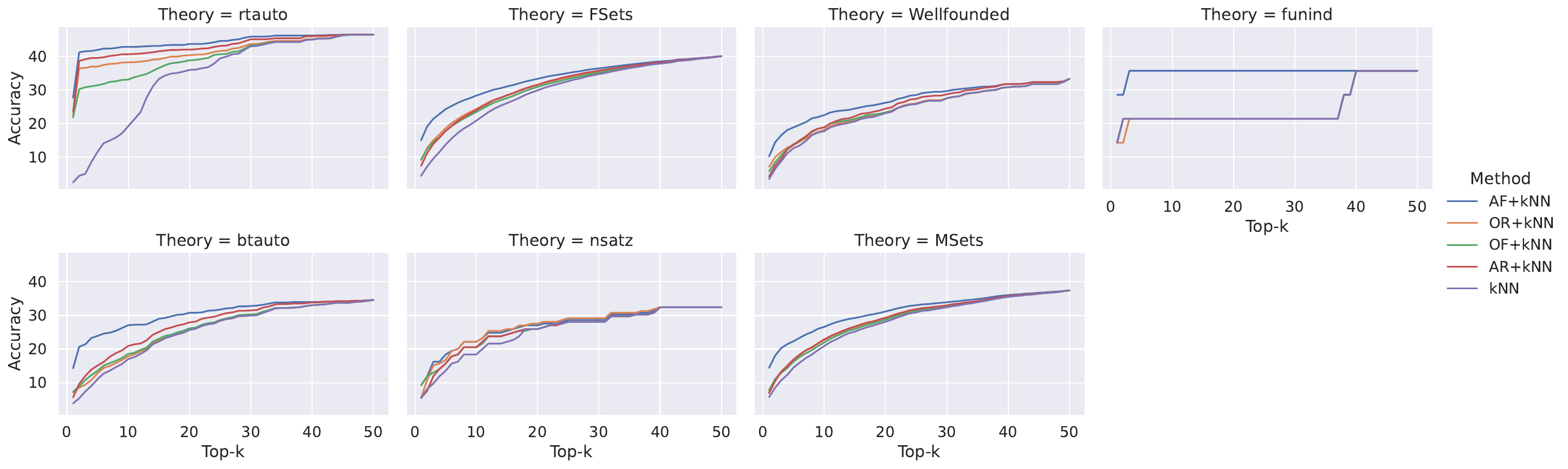}}
\caption{Top-$k$ accuracies in the test theories. It denotes how often the label is predicted in the first $k$ predictions. The symbol \textit{+} denotes using the rules learned by a certain predicate category to reorder the preselections.}
\label{fig:acc}
\end{figure*}

According to parameter optimization, we choose the rules with the best parameter and test the performance in the test dataset.
Table~\ref{tab:test_f1} shows the F-1 scores in the test dataset.
Using a background knowledge consisting of AF predicate definitions during training results in rules which perform best during testing.
This owes to that AF can learn precise rules to characterize the usage of tactics.
In comparison, the rules learned by AR are too general, and the rules learned by OF and OR are too specific.
In all theories, except those consisting of only a few proof states (\texttt{funind} has only 14 proof states), training with OF, OR, and AR background knowledge results in rules performing well on the test data F-1 scores.
We also evaluated whether the combination of ILP and $k$-NN can improve the accuracy of $k$-NN.
The algorithm of reordering is explained in Section~\ref{sec:reorder}.
Figure~\ref{fig:acc} shows the results of the top-$k$ accuracies in different theories.
In all the theories, the combination of ILP and $k$-NN increases the accuracies of $k$-NN.

\section{Case Studies and Limitations}
To illustrate that we indeed learn precise rules, besides the example of \texttt{simpl} presented in Section~\ref{sec:intro}, we present three more examples in this section.
The rule of \texttt{trivial} suits the goal $A \rightarrow B = B$.
First, \texttt{trivial} introduces $A$ as a hypothesis, changing the proof state to 
$H: A \vdash B = B$. Next, \texttt{trivial} can automatically prove $B = B$.
The rule of \texttt{auto} aligns the proof state of the structure $H: B \vdash A \lor B$.
The tactic \texttt{auto} decomposes the disjunction, and the goal changes to either proving $A$ or proving $B$. Then, it proves $B$ with the hypothesis.
In contrast, \texttt{trivial} cannot decompose the disjunction.
The rule of \texttt{intuition} suits the goal $A \leftrightarrow A$ which cannot be proved by \texttt{auto}.
In comparison, \texttt{intuition} can perform stronger automation than \texttt{auto} and can prove it.

\begin{minted}{prolog}
tac(A,"trivial") :-
  goal_node(prod,A,B,C),goal_node(const,A,D,E), goal_above(A,B,D),
  goal_node(const,A,F,E),goal_above(A,B,F),eq_goal_term(A,F,D).
tac(A,"auto") :-  
  goal_node(coq_Init_Logic_or,A,B,C),goal_node(const,A,D,E),
  goal_above(A,B,D),hyp_node(const,A,F,G,E),eq_goal_hyp_term(A,D,G).    
tac(A,"intuition") :-
  goal_node(coq_Init_Logic_iff,A,B,C),goal_node(const,A,D,E),
  goal_above(A,B,D),goal_node(const,A,F,E),eq_goal_term(A,F,D)
\end{minted}

Albeit we can learn several reasonable rules, many tactics are difficult to describe. 
There are several reasons for the difficulties.
First, our current work cannot generalize tactics with different arguments.
For instance, assume there are two tactics \texttt{apply H1} and \texttt{apply H2} where \texttt{H1} and \texttt{H2} are names of hypotheses.
They are regarded as different tactics but may have the same behavior.
Second, the usage of some tactics such as \texttt{induction} is inherently complicated~\cite{nagashima2019lifter}.
Third, the same mathematical theorem can proved in various ways which leads to many overlaps between the usage of tactics.

\section{Related Work}
There are several tasks of machine learning for theorem proving.
\textit{Premise selection} is probably the most well-discovered task. 
It studies the question of how to predict possibly useful lemmas for a given theorem.
Quite a lot of classical learning methods~\cite{alama2014premise,gauthier2015premise} and neural networks~\cite{irving2016deepmath} have been applied to premise selection. 
The most relevant task to our work is learning-based formal theory proving.
Researchers have investigated both employing machine learning to learn from human-written proofs~\cite{gauthier2021tactictoe} and guide some sophisticated software to automatically construct proofs~\cite{kaliszyk2018reinforcement}.




\section{Conclusion and Future Work}
We have developed the first application of ILP to interactive theorem proving. For this, we have developed new feature predicates, able to dynamically calculate features based on the representation of AST of the proof state. We proposed a method for using ILP effectively for tactic prediction.
We experimentally evaluated the rules learned by ILP and compared them to practically used prediction mechanisms in ITPs. The experiments confirm 
that the method gives explainable tactic predictions. 
Our work shows the potential of applications of ILP to improve ITP tactic suggestion methods.

Several improvements are possible.
We would like to use our work with stronger ILP systems, such as Popper. 
 However, given that our background knowledge includes predicates with high arity and our method builds large rules with many variables, the underlying ASP (SAT) solver used by Popper struggles with the generation of models. Improvements to our encoding and recent work on improving the performance of Popper can make this research direction viable in the near future.
Next, it is interesting to use ILP to capture the relations between arguments of tactics and the objects to which the arguments refer.
Finally, we plan to investigate the application of ILP to other ITP tasks~\cite{zhang2023learning}.

%
%

\begin{credits}
\subsubsection{\ackname}
This work was supported by the Czech Science Foundation Grant No. 22-06414L, the Cost action CA20111 EuroProofNet, the ERC PoC project \textit{FormalWeb3} no. 101156734, Amazon Research Awards, the EU ICT-48 2020 project no. 952215 TAILOR, the ERC CZ project no. LL1902 POSTMAN, and the
University of Innsbruck doctoral scholarship \textit{promotion of young talent}. 

\end{credits}

\bibliographystyle{splncs04}
\bibliography{mybibliography}

\end{document}